\documentclass[a4paper,11pt]{article}
\pdfoutput=1 

\usepackage{jinstpub} 

\title{\boldmath The calorimeter of the Mu2e experiment at Fermilab}

\author[a]{N.~Atanov,}
\author[a]{V.~Baranov,}
\author[a]{J.~Budagov,}
\author[e]{F.~Cervelli,}
\author[b]{F.~Colao,}
\author[b]{M.~Cordelli,}
\author[b]{G.~Corradi,}
\author[b]{E.~Dan\'e,}
\author[a]{Yu.I.~Davydov,}
\author[e,1]{S.~Di Falco,\note{Corresponding author.}}
\author[b,j]{E.~Diociaiuti,}
\author[e,g]{S.~Donati,}
\author[b,k]{R.~Donghia,}
\author[c]{B.~Echenard,}
\author[c]{K.~Flood,}
\author[b]{S.~Giovannella,}
\author[a]{V.~Glagolev,}
\author[i]{F.~Grancagnolo,}
\author[b]{F.~Happacher,}
\author[c]{D.G.~Hitlin,}
\author[b,d]{M.~Martini,}
\author[b]{S.~Miscetti,}
\author[c]{T.~Miyashita,}
\author[e,f]{L.~Morescalchi,}
\author[h]{P.~Murat,}
\author[e]{G.~Pezzullo,}
\author[c]{F.~Porter,}
\author[e]{F.~Raffaelli,}
\author[e]{T.~Radicioni,}
\author[b,d]{M.~Ricci,}
\author[b]{A.~Saputi,}
\author[b]{I.~Sarra,}
\author[e]{F.~Spinella,}
\author[i]{G.~Tassielli,}
\author[a]{V.~Tereshchenko,}
\author[a]{Z.~Usubov,}
\author[c]{R.Y.~Zhu}

\affiliation[a]{Joint Institute for Nuclear Research, Dubna, Russia}
\affiliation[b]{Laboratori Nazionali di Frascati dell'INFN, Frascati, Italy}
\affiliation[c]{California Institute of Technology, Pasadena, United States}
\affiliation[d]{Universit\`a ``Guglielmo Marconi'', Roma, Italy}
\affiliation[e]{INFN Sezione di Pisa, Pisa, Italy}
\affiliation[f]{Dipartimento di Fisica dell'Universit\`a di Siena, Siena, Italy}
\affiliation[g]{Dipartimento di Fisica dell'Universit\`a di Pisa, Pisa,
 Italy}
\affiliation[h]{Fermi National Laboratory, Batavia, Illinois, USA}
\affiliation[i]{INFN Sezione di Lecce, Lecce, Italy}
\affiliation[j]{Dipartimento di Fisica dell'Universit\`a di Roma Tor Vergata, Rome, Italy}
\affiliation[k]{Dipartimento di Fisica dell'Universit\`a degli Studi Roma Tre, Rome, Italy}

\emailAdd{stefano.difalco@pi.infn.it}

\abstract{
The Mu2e experiment at Fermilab looks for Charged Lepton Flavor Violation (CLFV) improving by 4 orders of magnitude the current experimental sensitivity for the muon to electron conversion in a muonic atom. A positive signal could not be explained in the framework of the current Standard Model of particle interactions and therefore would be a clear indication of new physics.
In 3 years of data taking, Mu2e is expected to observe less than one background event mimicking the electron coming from muon conversion.
Achieving such a level of background suppression requires a deep knowledge of the experimental apparatus: a straw tube tracker, measuring the electron momentum and time, 
a cosmic ray veto system rejecting most of cosmic ray background and a pure CsI crystal calorimeter, that will measure time of flight, energy and impact position of the converted electron.
The calorimeter has to operate in a harsh radiation environment, in a 10$^{-4}$ Torr vacuum and inside a 1 T magnetic field. The results of the first qualification tests of the calorimeter components are reported together with the energy and time performances expected from the simulation and measured in beam tests of a small scale prototype.
}


\proceeding{14th Topical Seminar on Innovative Particle and Radiation Detectors
(IPRD16)\\ 3 - 6 October 2016\\   Siena, Italy}

\begin{document}
\maketitle
\flushbottom

\section{Introduction}
\label{sec:intro}

After the discovery of lepton flavor violation in neutrino oscillation, the search for Charged Lepton Flavor Violation (CLFV) is one of the most important activities in particle physics. In the Standard Model of particle interactions the occurrence of such a process is predicted to be extremely rare, far  below the possible experimental reach. On the other hand, many extensions of the Standard Model predict CLFV rates that may be observed by the next generation of experiments~\cite{degouvea}.

The Mu2e experiment~\cite{TDR} at Fermilab aims to observe the neutrinoless conversion of a muon into an electron in the field of an Aluminum atom. In this two-body process the energy of the emerging electron is fixed (104.967 MeV) and the possible sources of background can be very efficiently suppressed.

In 3 years of running, $\sim10^{20}$ protons will be delivered to Mu2e and $\sim10^{18}$ muons will be stopped in the Aluminum stopping target. This huge amount of data will allow for a factor $10^4$ improvement on the sensitivity to the ratio between the rate of the neutrinoless muon conversions into electrons and the rate of ordinary muon capture in Al nucleus:
\begin{equation}
R_{\mu e}=\frac{\mu^{-}+Al\rightarrow e^{-}+Al}{\mu^{-}+Al\rightarrow \nu_{\mu} + Mg} 
\end{equation}

Even in case that no signal is observed, Mu2e will achieve a remarkable result: a limit $ R_{\mu e}<6\times 10^{-17}$ at 90\% confidence level, that is $10^4$ times better than the current limit set by the Sindrum II experiment~\cite{sindrum2}.

\section{The Mu2e experiment}
\label{sec:mu2e}

\begin{figure}[htbp]
\centering
\includegraphics{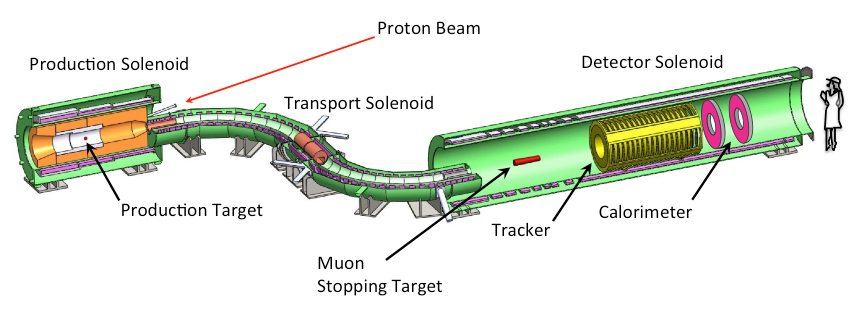}
\caption{\label{fig:mu2e} The Mu2e experiment.}
\end{figure}

The Mu2e experimental apparatus (figure~\ref{fig:mu2e}) consists of 3 superconducting solenoids: the production solenoid, where an 8 GeV proton beam is sent against a tungsten target and pions and kaons produced in the interactions are guided by a graded magnetic field towards the transport solenoid; a transport solenoid, with a characteristic 'S' shape, that transfers the negative particles with the desired momentum ($\sim$50 MeV) to the detector solenoid and absorbs most of the antiprotons thanks to a thin window of low Z material that separates the two halves of the solenoid; the detector solenoid, where the Aluminum muon stopping target is located and a graded field directs the electrons coming from the muon conversion to the tracker and the calorimeter. 

\begin{figure}[hb]
\centering
\includegraphics[height=4.cm]{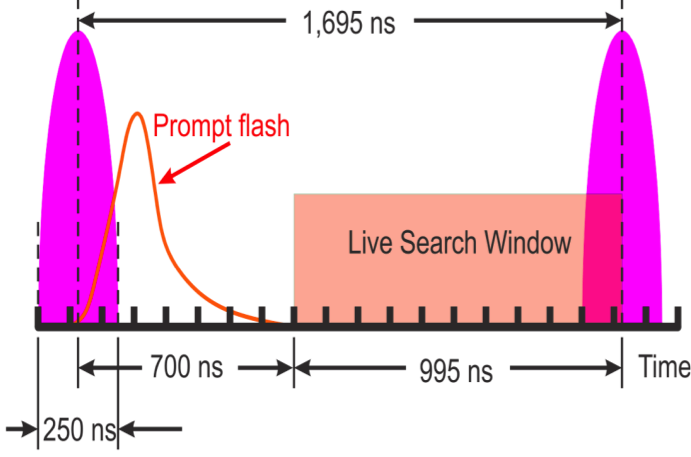}
\caption{\label{fig:pulsedbeam} The Proton bunches and the live search window used for data analysis.}
\end{figure}

The 8 GeV proton beam has the pulsed structure shown in figure~\ref{fig:pulsedbeam}. Each bunch lasts $\sim 250$ ns and contains $\sim 3\times 10^7$ protons.  The bunch period of $\sim 1.7\mu s$ facilitates exploitation of the time difference between the muonic Aluminum lifetime ($\tau = 864$ ns) and  the prompt backgrounds due to pion radiative decays, muon decays in flight and beam electrons, that are all concentrated within few tens of ns from the bunch arrival: a live search window delayed by 700 ns with respect to the bunch arrival suppresses these prompt backgrounds to a negligible level.

In order to achieve the necessary background suppression, it's important to have a fraction of protons out of bunch, or {\em extinction factor}, lower than $10^{-10}$. The current simulations of the accelerator optics predict an extinction factor better than required. The extinction factor will be continuously monitored by a dedicated detector located downstream of the production target.

\begin{figure}[htbp]
\centering 
\includegraphics[height=3cm]{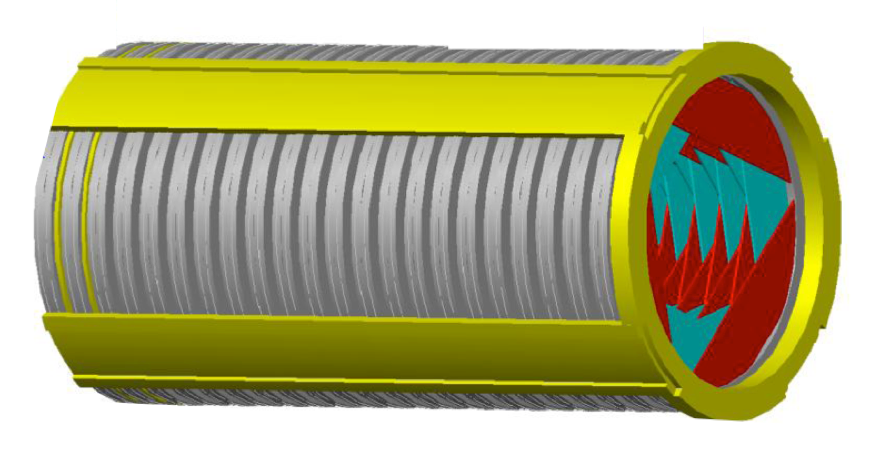}
\qquad
\includegraphics[height=3cm]{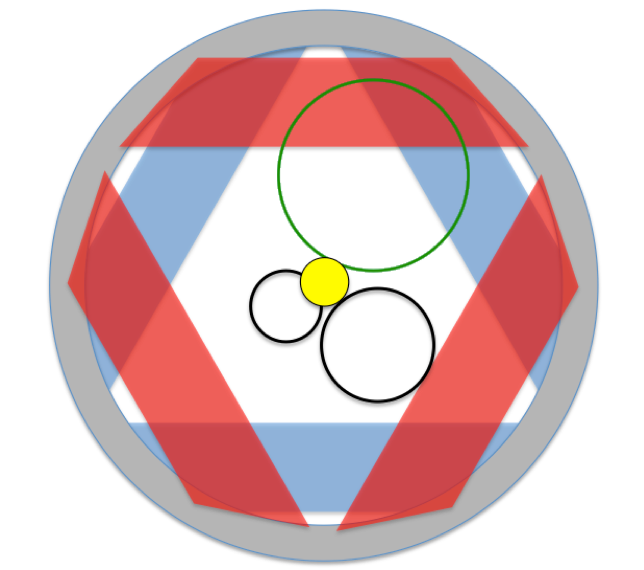}
\caption{\label{fig:tracker} Left: the Mu2e tracker with its 18 stations of straw tube panels. Right: section of the tracker showing how the large amount of low momentum particles (black circles) originated in the stopping target (yellow circle) doesn't cross the active detector volume.}
\end{figure}

The Mu2e Tracker consists of about 21000 low mass straw tubes oriented transverse to the solenoid axis and grouped into 18 measurement stations distributed over a distance of $\sim$3 m (figu-re \ref{fig:tracker}.left). Each straw tube is
instrumented on both sides with TDCs to measure the particle crossing time and ADCs to measure the specific energy loss dE/dX, that can be used to separate the electrons from highly ionizing particles. 
The central hole of radius $R\sim380$ mm precludes detection of charged particles with momentum lower than $\sim 50$ MeV/c (figure \ref{fig:tracker}.right).
The core of the momentum resolution for 105 MeV electrons is expected to be better than 180 keV/c, sufficient to suppress background electrons produced in the decays of muons captured by Aluminum nuclei.

\begin{figure}[htbp]
\centering 
\includegraphics[height=3cm]{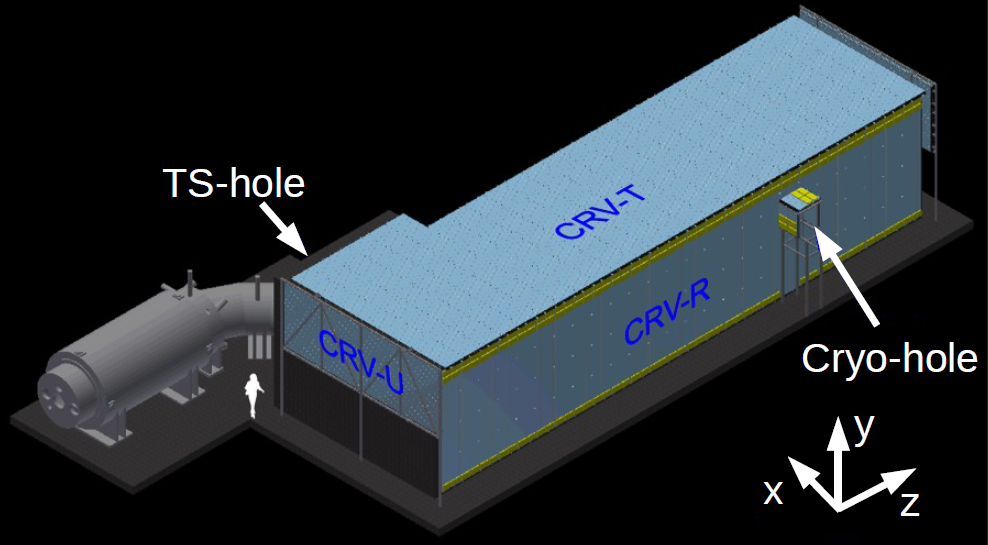}
\qquad
\includegraphics[height=3.5cm]{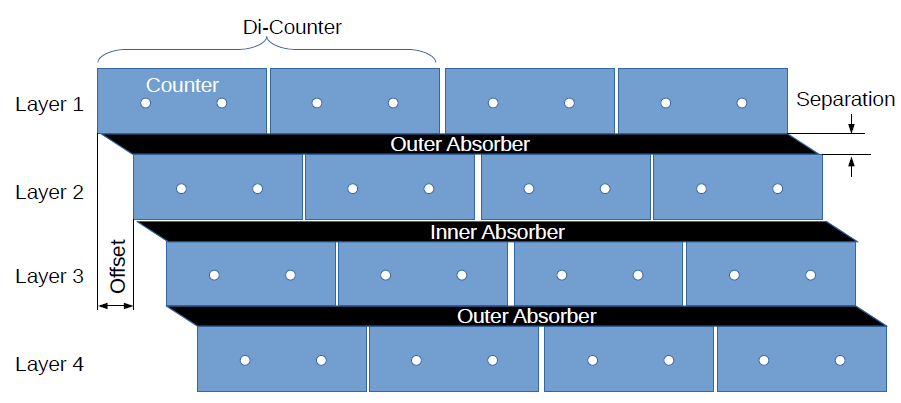}
\caption{\label{fig:crv} Left: the Mu2e cosmic ray veto system covering the detector solenoid and the last part of the transport solenoid. Right: the 4 layers of scintillators interleaved with Aluminum absorbers.}
\end{figure}

The background due to cosmic muons ($\delta$ rays, muon decays or misidentified muons) is suppressed by a cosmic ray veto system covering the whole detector solenoid and half of the transport solenoid (figure \ref{fig:crv}.left). The detector consists of  four layers of polystyrene scintillator counters interleaved with Aluminum absorbers (figure \ref{fig:crv}.right).  Each scintillator is read out via two embedded wavelength shifting fibers by silicon photomultipliers (SiPMs) located at each end. 
The veto is given by the coincidence of three out of four layers.
An overall veto efficiency of 99.99\% is expected. This corresponds to $\sim$ 1 background event in 3 years of data taking. An additional rejection factor will be provided by particle identification obtained by combining tracker and calorimeter information\footnote{An irreducible background of $\sim$0.1 electrons induced by cosmic muons in 3 years will nonetheless survive to particle identification~\cite{TDR}.}.

\section{The Mu2e electromagnetic calorimeter}
\label{sec:ecal}

The Mu2e electromagnetic calorimeter (ECAL) is needed to:
\begin{itemize}
\item identify the conversion electrons;
\item provide, together with tracker, particle identification to suppress muons and pions mimicking the conversion electrons; 
\item provide a standalone trigger to measure tracker trigger and track
reconstruction efficiency;
\item (optional) seed the tracker pattern recognition to reduce the number of possible hit combinations.
\end{itemize}

ECAL must operate in an harsh experimental environment:
\begin{itemize}
\item a magnetic field of 1 T;
\item a vacuum of $10^{-4}$ Torr;
\item a maximum ionizing dose of 100 krad for the hottest region at lower radius and $\sim$ 15 krad for the region at higher radius (integrated in 3 years including a safety factor of 3);
\item a maximum neutron fluence of $10^{12}\ n/cm^2$ (integrated in 3 years including a safety factor of 3)
\item a high particle flux also in the live search window.
\end{itemize}

\begin{figure}[htbp]
\centering
\includegraphics[height=6.cm]{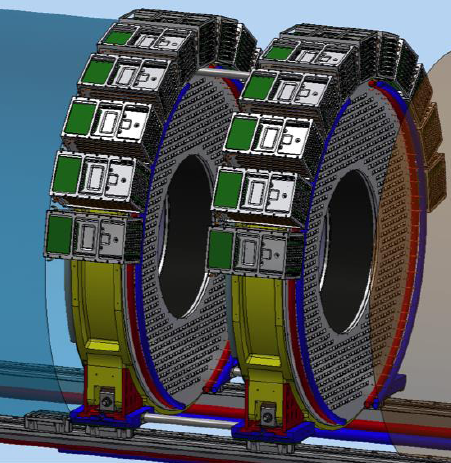}
\caption{\label{fig:ecal} The Mu2e electromagnetic calorimeter.}
\end{figure}
The solution adopted for the Mu2e calorimeter (figure \ref{fig:ecal}) consists of two annular disks of
undoped CsI crystals placed at a relative distance of $\sim 70$ cm, that is approximately half pitch of the conversion electron helix in the magnetic field. 
The disks have an inner radius of 37.4 cm and an outer radius of 66 cm. The design minimizes the number of low-energy particles that intersect the calorimeter while maintaining an high acceptance for the signal. 

Each disk contains 674 undoped CsI crystals of $20\times3.4\times3.4\ cm^3$. This granularity has been optimized taking into account the light collection for readout photosensors, the particles pileup, the time and energy resolution. 

Each crystal is read out by two arrays of UV-extended silicon photomuliplier sensors (SiPM). The SiPMs signal is amplified and shaped by the Front-End Electronics (FEE) located on their back. The voltage regulators and the digital electronics, used to digitize the signals, are located in crates disposed around the disks.

\subsection{CsI crystals}

The characteristics of the pure CsI crystals are reported in  table~\ref{tab:CsI}. These crystals have been preferred to the other candidates because of their emission frequency, well matching the sensitivity of commercial photosensors, their good time and energy resolution and their reasonable cost.

\begin{table}[htbp]
\centering
\caption{\label{tab:CsI} Characteristics of pure CsI crystals.}
\smallskip
\begin{tabular}{|l|c|}
\hline
~& CsI\\
\hline
density ($g/cm^3$) & 4.51\\
radiation length ($cm$) & 1.86\\
Moli\`ere radius ($cm$) & 3.57\\
interaction length ($cm$) & 39.3\\
dE/dX ($MeV/cm$) & 5.56\\
refractive index ($cm$) & 1.95\\
peak luminescence ($nm$) & 310\\
decay time ($ns$) & 26\\
light yield (rel. to NaI) & 3.6\%\\
variation with temperature & -1.4\%/$^{o}$C\\
\hline
\end{tabular}
\end{table}

Each crystal will be wrapped with 150 $\mu m$ of Tyvek 4173D.

Quality tests on a set of pure CsI crystals from SICCAS (China), Optomaterial (Italy) and ISMA (Ukraine) have been performed in Caltech and at the INFN Laboratori Nazionali di Frascati (LNF)\cite{lnfcsi}.
The results can be summarized as follows:
\begin{itemize}
\item a light yield of 100 p.e./MeV when measured with a 2'' UV extended EMI PMT;
\item an emission weighted longitudinal transmittance varying from 20\% to 50\% depending on the crystal surface quality;
\item a light response uniformity corresponding to a variation of $0.6\%/cm$;
\item a decay time $\tau\sim 30$ ns with, in some cases, a small slow component with $\tau\sim 1.6\mu s$;
\item a light output reduction lower than $40\%$ after an irradiation with a total ionizing dose of 100 krad;
\item a negligible light output reduction but a small worsening of longitudinal response uniformity after an irradiation with a total fluence of $9\times 10^{11} n/cm^2$;
\item a radiation induced readout noise in the Mu2e radiation environment equivalent to less than 600 KeV.
\end{itemize}

\subsection{Photosensors}

\begin{figure}[htbp]
\centering 
\includegraphics[height=4cm]{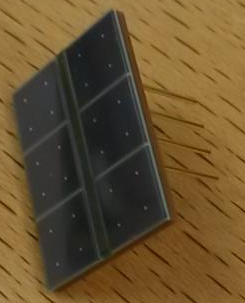}
\qquad
\includegraphics[height=4cm]{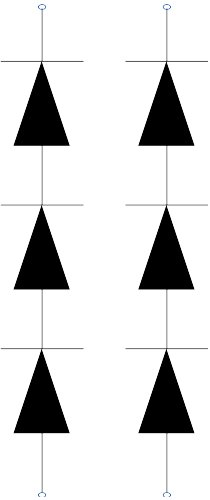}
\caption{\label{fig:sipmarr} Left: one SiPM array to be used in the Mu2e calorimeter. Right: the two series connections in the SiPM array.}
\end{figure}

Figure \ref{fig:sipmarr} shows one of the two SiPM arrays used to read each crystal.

The array is formed by two series of 3 SiPMs. The two series are connected in parallel by the Front End electronics to have a x2 redundancy.
The series connection reduces the global capacitance, improving the signal decay time to less than 100 ns. It also minimizes the output current and the power consumption. 

Each SiPM has an active surface of 6x6 $mm^2$ and is UV-extended with a photon detection efficiency (PDE) at the CsI emission peak ($\sim$315 nm) of $\sim$ 30\%.

Tests on single SiPM prototypes from different vendors (Hamamatsu, SENSL, Advansid) have been performed at LNF and INFN Pisa. 

The gain is better than $10^6$ at an operating voltage $V_{OP}=V_{BR}+3V$, where $V_{BR}$ is the breakdown voltage of the SiPM. When coupled in air with the CsI crystal the yield is $\sim$ 20 p.e./MeV.
The noise correspond to an additional energy resolution of $\sim$ 100 keV.

A test of neutron irradiation with a fluence of $4\times 10^{11}$ neutrons/$cm^2$ 1 MeV equivalent\footnote{Since SiPMs are partially shielded by the crystals, this corresponds for the SiPMs for the 3 years of Mu2e running to a safety factor of $\sim$2.} and a SiPM temperature kept stable at 25$^{o}C$, has produced a dark current increase from 60 $\mu A$ to 12 mA and a gain decrease of 50\%. 

A test with photon radiation corresponding to a total ionizing dose of 20 krad has produced negligible effects on gain and dark current.

In order to reduce the effects of radiation damage and to keep the power consumption at a reasonable level the  SiPM temperature will be kept stable at $0^{o}C$.

The qualification tests of the SiPM array preproduction are in progress at Caltech, LNF and INFN Pisa and will evaluate:
\begin{itemize}
\item the I-V characteristics of the single SiPMs and of each series;
\item the breakdown voltage $V_{BR}$ and the operating voltage $V_{OP}=V_{BR}+3V$  of the single SiPMs and of each series;
\item the absolute gain and the PDE relative to a reference sensor at  $V_{OP}$ for the single SiPMs and for each series;
\item the mean time to failure (MTTF) through an accelerated aging test at $55^{o}$;
\item the radiation damage due to neutron, photons and heavy ions.
\end{itemize}

\subsection{Read out electronics}

\begin{figure}[htbp]
\includegraphics[height=4cm]{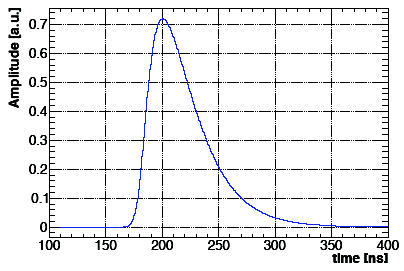}
\qquad
\includegraphics[height=4cm]{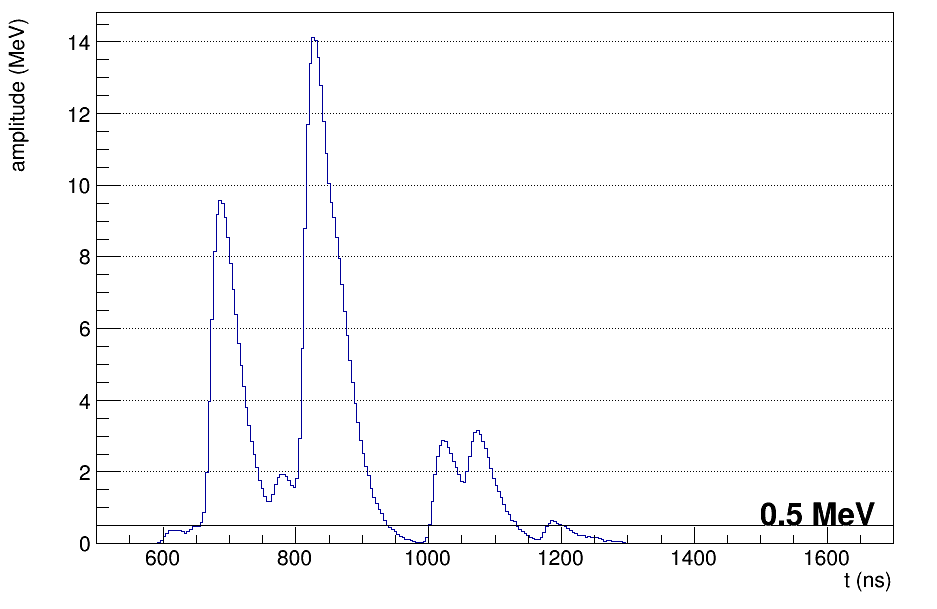}
\caption{\label{fig:signals} Left: example of signal produced by the shaper amplifier. Right: particles pileup observed with the waveform digitizer. The digitization threshold used for the zero suppression is also shown.}
\end{figure}

In the front end electronics board, directly connected to the SiPM array,
the signals coming from the two series are summed in parallel and then shaped and amplified in order to obtain a signal similar to the one shown in figure \ref{fig:signals}.left. This shaping aims to reduce the pileup of energy deposits due to different particles and to optimize the resolution on the particle arrival time.
 
The shaped signals are sent to a waveform digitizer boards where they are digitized at a sampling frequency of 200 MHz using a 12 bits ADC.

The most critical components of the waveform digitizer board are: the SM2150T-FC1152 Microsemi SmartFusion2 FPGA, the Texas Instrument ADS4229 ADC and the Linear Tecnologies LTM8033 DC/DC converter.

The FPGA is already qualified by the vendor as SEL and SEU free and will be tested only together with the assembled board.

The DC/DC converter, tested in a 1 T magnetic field, still maintain an efficiency of $\sim$ 65\%. Negligible effects on output voltage and efficiency have been observed after neutron and photon irradiation corresponding to 3 years of Mu2e running.

The ADCs have also been irradiated with neutrons and photons equivalent to  3 years of Mu2e running and have shown no bit flips or loss of data.

\subsection{Energy and time calibration}

The energy and time calibration of the Mu2e calorimeter could be performed using different calibration sources~\cite{nimecal}.

A 6 MeV activated liquid source (Fluorinert)~\cite{babar} can be circulated into pipes located in front of each disk to set the absolute energy scale.

A laser calibration system can be used to pulse each crystal to equalize the time offset and the energy response of each SiPM array channel.

Also minimum ionizing cosmic muons can be used  to equalize the time offset and the energy response.

The energy-momentum matching for electrons produced by  muons decaying in the orbit of the Al atom or by pion two body decays can be used to set the energy scale and to determine the time offset with respect to the tracker. These low momentum electrons mostly pass through the hole of the calorimeter disks but can be used in special calibration runs with reduced magnetic field.

\section{Beam test of a small matrix}

A  calorimeter prototype consisting of a 3$\times$3 matrix of $3\times3\times20$ cm$^3$ undoped CsI crystals wrapped in 150 $\mu m$ of Tyvek R and read by one 12$\times$12 mm$^2$ SPL TSV SiPM by Hamamatsu has been tested with an electron beam at the Beam Test Facility (BTF) in Frascati during April 2015.

The results obtained are coherent with the ones predicted by the simulation:
\begin{itemize}
\item a time resolution better than 150 ps for 100 MeV electrons;
\item an energy resolution of $\sim$ 7\% for 100 MeV electrons with a $50^o$ incidence angle\footnote{This is the most probable incidence angle for conversion electrons reaching the Mu2e calorimeter. The values range from 40$^o$ to 60$^o$.}, dominated by the energy leakage due to the few Moli\`ere radii of the prototype.
\end{itemize}

\section{Calorimeter performances predicted by simulation}

A detailed Monte Carlo simulation has been developed to optimize the calorimeter design.

The simulation corresponding to the final design predicts the following performances for 100 MeV electrons:
\begin{itemize}
\item a time resolution of $\sim$110 ps;
\item an energy resolution of $\sim$4\%;
\item a position resolution of 1.6 cm in both the transverse coordinates.
\end{itemize}

Combining the calorimeter and tracker time and energy/momentum information it's possible to distinguish between 100 MeV electrons and  muons with the same momentum: an electron efficiency of 94\% and  a corresponding muon rejection factor of 200 have been obtained.

The calorimeter information can be used to seed the track reconstruction improving the reconstruction efficiency and contributing to remove the background due to tracks with poorly reconstructed momentum.

A standalone software trigger based on the calorimeter information only is able to achieve an efficiency of 60\% on conversion electrons while suppressing the background trigger rate by a factor 400.

A combined software trigger using both tracker and calorimeter information is able to achieve an efficiency of 95\% on conversion electrons with a background rejection factor of 200.

\section{Conclusions and outlook}
The Mu2e calorimeter is a key component of the Mu2e experiment.

The calorimeter design is now mature:
quality tests have shown that the chosen components are able to operate in the Mu2e harsh environment. 

Monte  Carlo simulation, supported by test beam results, shows that the 
current design meets the requirements on muon identification, seeding of track reconstruction and trigger selection. 
 
A beam test of a larger scale prototype with 50 pure CsI crystals read by 100 SiPM arrays will be performed in the next months.

\acknowledgments
This work was supported by the EU Horizon 2020 Research and Innovation Programme under the Marie Sklodowska-Curie Grant Agreement No. 690835.

\end{document}